\documentclass[
    ,final            
,numberedheadings 
  ]
  {aipproc}

\layoutstyle{8x11single}

\usepackage{amssymb,amsmath,mathrsfs,amsthm}

\def\im{\mathrm{Im\,}}
\def\ad{\mathrm{ad\,}}

\usepackage{eucal}

\def\im{\mathrm{Im\,}}
\def\ad{\mathrm{ad\,}}

\def\openone{\leavevmode\hbox{\small1\kern-3.3pt\normalsize1}}

\def\bbbz{\Bbb{Z}}

\def\ad{\mbox{ad\,}}

\def\im{\mbox{Im\,}}

\def\diag{\mbox{diag\,}}

\def\bbbr{{\Bbb R}}

\def\bbbz{{\Bbb Z}}
\def\openone{\leavevmode\hbox{\small1\kern-3.3pt\normalsize1}}

\allowdisplaybreaks
\newtheorem{lemma}{Lemma}
\newtheorem{theorem}{Theorem}

\usepackage{epsf}

\def\newpic#1{%
   \def\emline##1##2##3##4##5##6{%
      \put(##1,##2){\special{em:point #1##3}}%
      \put(##4,##5){\special{em:point #1##6}}%
      \special{em:line #1##3,#1##6}}}

\newpic{}

\def\bbbr{{\Bbb R}}

\def\bbbz{{\Bbb Z}}

\def\ad{\mbox{ad}\,}

\def\im{{\rm Im}\,}

\def\openone{\leavevmode\hbox{\small1\kern-3.3pt\normalsize1}}

\begin{document}

\title{On  new types of integrable 4-wave interactions }


\author{ Vladimir S. Gerdjikov  }{
  address={Institute for Nuclear Research and Nuclear
Energy, Bulgarian Academy of Sciences\\
72 Tsarigradsko chaussee, 1784 Sofia, Bulgaria }
}

\begin{abstract}

We start with a Riemann-Hilbert Problems (RHP) with canonical normalization
whose sewing functions depends on two or more  additional variables.
Using Zakharov-Shabat theorem we are able to construct a family of
ordinary differential operators for which the solution of the RHP is a
common fundamental analytic solution. This family of operators obviously
commute provided their coefficients satisfy certain nonlinear evolution equations.
Thus we are able to construct new classes of integrable
nonlinear evolution equations.
We illustrate the method with an example of a new type 4-wave interactions. Its
Lax pair consists of  operators which are both quadratic in the
spectral parameter $\lambda$ and take values in the $so(5)$ algebra.
\end{abstract}

\maketitle


  \section{ Introduction}
The $N$-wave interactions  describe special types of wave-wave interactions \cite{Za1*76,CaDem,ZMNP,GVYa*08,DokLeb,KRB,ISK*99} which play
important role in various fields in physics.

The topic quickly attracted  mathematicians from spectral theory,
dynamical systems, Lie algebras, Hamiltonian dynamics,
differential geometry, see \cite{ZMNP,MaZa*79,Za*80,Za1*80,Fer*95,GVYa*08,DokLeb}
and the numerous references therein.
It attracted also a number of physicists because they found
important applications  of these nonlinear evolution equations (NLEE)
in fluid mechanics, nonlinear optics, superconductivity, plasma physics etc.
As a result many different approaches for investigating the soliton
equations and constructing their Lax representations,
soliton solutions, integrals of motion, Hamiltonian hierarchies etc. were developed, see
\cite{ZM1,MaZa*79,Corn*78,ZMNP,GK*81,KRB,Za1*80,DJK1*80}.

In the present paper, which is a natural extension of \cite{107s}, we propose an alternative approach to the same class of
equations using as a starting point  the Riemann-Hilbert problem (RHP)
\cite{ZaSh*74a,ZaSh*79,Za*80,Za1*80,ZMNP,ZaMaLomi,DJK2*80}; the importance of the canonical normalization
of RHP was noticed in \cite{21,18}. Our aim is to show that this allows
one to construct rings of commuting operators and in addition gives a tool to
study their spectral properties.

In Section 2 below we start with some preliminaries concerning the RHP and their reductions.
 In Section 3 we use jets of order 1 to reproduce well known results
about the $4$-wave equations, see Chapter 3 of \cite{ZMNP}.
In Section 4 we use jets of order 2 which allows us to construct
new types of integrable $4$-wave interactions whose interaction terms contain
quadratic and cubic nonlinearities, as well as $x$-derivatives.
The last Section contains discussion and conclusions.

\section{Preliminaries}
\subsection{ RHP with canonical normalization}
Let us formulate the RHP:
\begin{equation}\label{eq:rhp}\begin{split}
 \xi^+( x ,t,\lambda) &=  \xi^-( x ,t,\lambda)  G( x ,t,\lambda) , \qquad \lambda^k\in\bbbr, \qquad
 \lim_{\lambda\to\infty} \xi^+( x ,t,\lambda) =\openone ,
 \end{split}\end{equation}
where $\xi^\pm( x ,t,\lambda) $ take values in the simple Lie group  $\mathfrak{G}$ with
Lie algebra $\mathfrak{g}$. $ \xi^+( x ,t,\lambda)$ (resp. $ \xi^-( x ,t,\lambda)$) is
an analytic functions of $\lambda$ for $\im \lambda^k >0$ (resp. for $\im \lambda^k <0$).
 For simplicity we consider particular type of dependence of the sewing
function $G( x ,t,\lambda) $ on the  auxiliary variables:
\begin{equation}\label{eq:Gxt}\begin{split}
i \frac{\partial G}{ \partial x  } -\lambda^k [J ,  G( x ,t,\lambda) ] &=0, \qquad
  i \frac{\partial G}{ \partial t } -\lambda^k [K,  G( x ,t,\lambda) ] =0.
\end{split}\end{equation}
where $k \geq 1$ is a fixed integer and $J,K $ are linearly independent elements of the Cartan subalgebra
$J,K  \in \mathfrak{h}\subset \mathfrak{g}$.

The canonical normalization of the RHP means that we can introduce the asymptotic expansion
\begin{equation}\label{eq:xi-as}\begin{split}
 \xi^\pm( x ,t,\lambda) = \exp Q( x ,t,\lambda), \qquad Q( x ,t,\lambda)=\sum_{k=1}^{\infty} Q_k( x ,t) \lambda^{-k} .
\end{split}\end{equation}
Since $\xi^\pm( x ,t,\lambda) $ are group elements then all $Q_k( x ,t)\in \mathfrak{g}$. However,
\begin{equation}\label{eq:33w}\begin{split}
 \mathcal{J} ( x ,t,\lambda)= \xi^\pm ( x ,t,\lambda) J  \hat{\xi}^\pm( x ,t,\lambda),
 \qquad  \mathcal{K}( x ,t,\lambda)=\xi^\pm ( x ,t,\lambda) K \hat{\xi}^\pm( x ,t,\lambda),
\end{split}\end{equation}
belong to the algebra $\mathfrak{g}$ for any $J$ and $K$ from $\mathfrak{g}$. Since
$K$ also belongs to the Cartan subalgebra $\mathfrak{h}$, then
\begin{equation}\label{eq:jets}\begin{split}
[\mathcal{J} ( x ,t,\lambda), \mathcal{K}( x ,t,\lambda)] =0.
\end{split}\end{equation}

An important tool in our considerations plays the well known  Zakharov-Shabat theorem \cite{ZaSh*74a,ZaSh*79} formulated below
\begin{theorem}\label{thm:ZSh}
Let $\xi^\pm (x,t,\lambda)$ be solutions to the RHP (\ref{eq:rhp}) whose sewing function depends on
the auxiliary variables $ x $ and $t$ via eq. (\ref{eq:Gxt}). Then $\xi^\pm (x,t,\lambda)$ are fundamental solutions
of the following set of differential operators:
\begin{equation}\label{eq:xi-L}\begin{split}
L \xi^\pm\equiv & i \frac{\partial \xi^\pm }{ \partial x  } + U ( x ,t,\lambda) \xi^\pm ( x ,t,\lambda) -
\lambda^k [J ,\xi^\pm( x ,t,\lambda)]=0, \\
M\xi^\pm\equiv & i \frac{\partial \xi^\pm }{ \partial t } + V( x ,t,\lambda) \xi^\pm( x ,t,\lambda) -\lambda^k [K,\xi^\pm( x ,t,\lambda)]=0.
\end{split}\end{equation}

\end{theorem}

\begin{proof}
The proof follows the lines of \cite{ZaSh*74a,ZaSh*79}. We introduce the functions:
\begin{equation}\label{eq:gpm}\begin{split}
g ^\pm( x ,t,\lambda) &= i \frac{\partial \xi^\pm }{ \partial x  } \hat{\xi}^\pm( x ,t,\lambda) +\lambda^k
\xi^\pm( x ,t,\lambda) J  \hat{\xi}^\pm( x ,t,\lambda), \\
k^\pm( x ,t,\lambda) &= i \frac{\partial \xi^\pm }{ \partial t } \hat{\xi}^\pm( x ,t,\lambda) +\lambda^k
\xi^\pm( x ,t,\lambda) K \hat{\xi}^\pm( x ,t,\lambda),
\end{split}\end{equation}
where $\hat{\xi}^\pm \equiv (\xi^\pm)^{-1}$,  and using (\ref{eq:Gxt}) prove that
\begin{equation}\label{eq:3gs}\begin{split}
g ^+( x ,t,\lambda) = g ^-( x ,t,\lambda), \qquad k^+( x ,t,\lambda) = k^-( x ,t,\lambda),
\end{split}\end{equation}
which means that these functions are analytic functions of $\lambda$ in the whole complex $\lambda$-plane.
Next we find that:
\begin{equation}\label{eq:g-as}\begin{split}
\lim_{\lambda\to\infty} g ^+( x ,t,\lambda) = \lambda^k J , \qquad
\lim_{\lambda\to\infty} k^+( x ,t,\lambda) = \lambda^k K.
\end{split}\end{equation}
and make use of Liouville theorem to get
\begin{equation}\label{eq:UV}\begin{split}
g ^+( x ,t,\lambda)& = g ^-( x ,t,\lambda)= \lambda^k J  - \sum_{l=1}^{k} U_{l} ( x ,t) \lambda^{k-l}, \\
k^+( x ,t,\lambda)& = k^-( x ,t,\lambda) =\lambda^k K - \sum_{l=1}^{k} V_{l} ( x ,t) \lambda^{k-l}.
\end{split}\end{equation}
We shall see below that the coefficients $U_{l} ( x ,t)$ and $V_{l} ( x ,t)$ can be expressed
in terms of the asymptotic coefficients $Q $ in eq. (\ref{eq:xi-as}).
\end{proof}

\begin{lemma}\label{lem:1}
The  operators $L $ and $M$ commute:
\begin{equation}\label{eq:LsM}\begin{split}
i \frac{\partial  U }{ \partial t } &- i \frac{\partial  V }{ \partial x  }
+ [  U ( x ,t,\lambda) - \lambda^k J ,  V( x ,t,\lambda) -\lambda^k K]=0.
\end{split}\end{equation}
where
\begin{equation}\label{eq:UV'}\begin{split}
U ( x ,t,\lambda) =\sum_{l=1}^{k} U_{ l } ( x ,t) \lambda^{k-l}, \qquad
V( x ,t,\lambda) = \sum_{l=1}^{k} V_{l} ( x ,t) \lambda^{k-l}.
\end{split}\end{equation}

\end{lemma}

\begin{proof}
The   operators $L $ and $M$ (\ref{eq:xi-L}) have  common fundamental analytic solutions (FAS), i.e. they must commute.
The eqs. (\ref{eq:LsM}) are an immediate consequence of (\ref{eq:xi-L}).
\end{proof}

\subsection{Jets of order $k$}

In what follows we will consider the jets of order $k$ of $\mathcal{J}(x,\lambda)$ and $\mathcal{K}(x,\lambda)$, see
(\ref{eq:jets}). We introduce them by:
\begin{equation}\label{eq:Jet1}\begin{split}
 \mathcal{J} ( x ,t,\lambda) &\equiv \left( \lambda^k \xi^\pm ( x ,t,\lambda) J_l \hat{\xi}^\pm( x ,t,\lambda) \right)_+
 = \lambda^k J  -  U  ( x ,t, \lambda), \\
 \mathcal{K}( x ,t,\lambda) &\equiv \left( \lambda^k \xi^\pm ( x ,t,\lambda) K \hat{\xi}^\pm( x ,t,\lambda) \right)_+
 = \lambda^k K -  V ( x ,t, \lambda).
\end{split}\end{equation}
The subscript $+$ used above means that we insert the asymptotic expansions of $\xi^\pm$ and their inverse (\ref{eq:xi-as}) and
cut off the terms with negative powers of $\lambda$.

Obviously $U (x) \in \mathfrak{g}$ can be expressed in terms of $Q (x)$. In doing this we take into
account (\ref{eq:jets}) and obtain  \cite{Helg}
\begin{equation}\label{eq:xi-ex}\begin{split}
\mathcal{J} ( x ,t,\lambda) = J  + \sum^{\infty}_{k=1} \frac{1}{k!} \ad_{Q}^k J ,  \qquad \mathcal{K}( x ,t,\lambda) =
K + \sum^{\infty}_{k=1} \frac{1}{k!} \ad_{Q}^k K,
\end{split}\end{equation}
and therefore for $U_{ l }$ we get:
\begin{equation}\label{eq:J_k}\begin{split}
U_{1}( x ,t) & = -\ad_{Q_1}J , \qquad  U_{2}( x ,t)  = -\ad_{Q_2}J  - \frac{1}{2}\ad_{Q_1}^2 J  \\
U_{3}( x ,t) & = -\ad_{Q_3}J  - \frac{1}{2}\left( \ad_{Q_2}\ad_{Q_1} +\ad_{Q_1}\ad_{Q_2}\right) J  - \frac{1}{6} \ad_{Q_1}^3J  \\
  & \vdots \\
U_{k}( x ,t) & = -\ad_{Q_k}J  - \frac{1}{2} \sum_{s+p=k}^{}\ad_{Q }\ad_{Q_p}  J
 - \frac{1}{6} \sum_{s+p+r=k}^{}\ad_{Q }\ad_{Q_p}\ad_{Q_r}  J   - \cdots - \frac{1}{k!} \ad_{Q_1}^k J  ,
\end{split}\end{equation}
and similar expressions for $V_l( x , t) $ with $J $ replaced by $K$.

\subsection{Reductions of polynomial bundles}

An important tool to construct new integrable NLEE is based on Mikhailov's group of reductions
\cite{Mik1}. Below we will use mainly $\bbbz_2$ and  $\bbbz_N$ with $N>2$ reduction groups.
The basic $\bbbz_2$-examples are as follows:
\begin{equation}\label{eq:T10.1}\begin{aligned}
&\mbox{a)} &\quad A \xi^{+,\dag} (x,t,\epsilon \lambda^*) \hat{A} &= \hat{\xi}^-(x,t,\lambda), &\quad
AQ^\dag (x,t,\epsilon\lambda^*) \hat{A} &=-Q(x,t,\lambda), \\
&\mbox{b)} &\quad B \xi^{+,*} (x,t,\epsilon \lambda^*) \hat{B} &= \xi^-(x,t,\lambda), &\quad
BQ^* (x,t,\epsilon\lambda^*) \hat{B} &=Q(x,t,\lambda), \\
&\mbox{c)} &\quad C \xi^{+,T} (x,t, -\lambda) \hat{C} &= \hat{\xi}^-(x,t,\lambda), &\quad
CQ^\dag (x,t,-\lambda) \hat{C} &=-Q(x,t,\lambda),
\end{aligned}\end{equation}
where $\epsilon^2 =1$ and $A$, $B$ and $C$ are elements of the group $\mathfrak{G}$ such that
$A^2=B^2=C^2=\openone$. As for the $\bbbz_N$-reductions we may have:
\begin{equation}\label{eq:T10.2}\begin{aligned}
D \xi^{\pm} (x,t, \omega \lambda) \hat{D} &= \xi^\pm(x,t,\lambda), &\qquad
DQ (x,t, \omega\lambda) \hat{D} &=Q(x,t,\lambda),
\end{aligned}\end{equation}
where $\omega^N=1$ and $D^N=\openone$.

These relations allow us to introduce algebraic relations between the matrix elements of
$Q(x,t,\lambda)$ which will be automatically compatible with the NLEE. The classes of inequivalent reductions of the $N$-wave
equations related to the low-rank simple Lie algebras are given in \cite{Joro,GGK*01,GIK,GKKV*08,GKKV*08,GKV*07,GV*V06}.

\section{The classical  $4$-wave interactions }

The 4-wave interactions were discovered by Zakharov (see Chapter 3 of \cite{ZMNP}) who used a $4\times 4$ Lax pair with an
additional reduction which effectively reduced it to the subalgebra $sp(4)$. It is well known that $sp(4)$
is isomorphic to  $so(5)$, so it is a matter of taste which one to choose. We remind that the root system $\Delta$ of
$so(5)$ has 8  roots (see \cite{Helg}):
\begin{equation}\label{eq:Dep}\begin{split}
\Delta \equiv \Delta_+\cup \Delta_-, \qquad \Delta_\pm =\{ \pm \alpha_1, \pm\alpha_2, \pm(\alpha_1+\alpha_2),
\pm(\alpha_1 +2\alpha_2)\},
\end{split}\end{equation}
where $\alpha_1=e_1-e_2$ and $\alpha_2=e_2$ are the two simple roots.

With this notations the Lax pair proposed by Zakharov takes the usual form for the $N$-wave equations
\begin{equation}\label{eq:1}\begin{split}
L\psi &= i \frac{\partial \psi}{ \partial x } + (U_1(x,t) + \lambda  J) \psi(x,t,\lambda)=0, \\
M\psi &= i \frac{\partial \psi}{ \partial t } + (V_1(x,t) + \lambda  K) \psi(x,t,\lambda)=0,
\end{split}\end{equation}
where the potentials $U_1$ and  $V_1$ are of the form:
\begin{equation}\label{eq:}\begin{aligned}
U_1(x,t) &= [J, Q_1(x,t)], &\qquad V_1(x,t) &= [K, Q_1(x,t)], \\
Q_1(x,t) &=\sum_{\alpha\in\Delta_+}^{} (q_\alpha(x,t) E_\alpha + p_\alpha(x,t) E_{-\alpha} )
\end{aligned}\end{equation}

We impose the natural reduction (\ref{eq:T10.1}a) which provide $p_\alpha=\epsilon q_\alpha^*$.
Thus the 4-wave equations take the form
\begin{equation}\label{eq:4w1}\begin{split}
i \left[ J , \frac{\partial Q_1}{ \partial t } \right] - i \left[ K , \frac{\partial Q_1}{ \partial x } \right]
+ \left[ [J, Q_1], [K,Q_1(x,t)]\right] =0,
\end{split}\end{equation}
or in components:
\begin{equation}\label{eq:4w2}\begin{split}
2i(a_1-a_2) \frac{\partial u_1}{ \partial t } &- 2i(b_1-b_2) \frac{\partial u_1}{ \partial x } +2\kappa v_2u_4 =0, \\
2ia_2 \frac{\partial u_2}{ \partial t } &- 2ib_2 \frac{\partial u_2}{ \partial x } +2\kappa (v_4u_3+u_4v_1) =0, \\
2ia_1 \frac{\partial u_4}{ \partial t } &- 2ib_1 \frac{\partial u_4}{ \partial x } +2\kappa (v_2u_3-u_2u_1) =0, \\
2i(a_1+a_2) \frac{\partial u_3}{ \partial t } &- 2i(b_1+b_2) \frac{\partial u_3}{ \partial x } -2\kappa u_2u_4 =0,
\end{split}\end{equation}
where $u_1 =q_{\alpha_1}$, $u_2 =q_{\alpha_2}$, $u_4 =q_{\alpha_1+\alpha_2}$ and  $u_3 =q_{\alpha_1+2\alpha_2}$.

\section{New types of $4$-wave interactions -- an example}

Here we will give an example of a new type of integrable 4-wave equations.

The Lax pair for these new equations will be provided by:
\begin{equation}\label{eq:1n}\begin{split}
L\psi &= i \frac{\partial \psi}{ \partial x } + (U_2(x,t) + \lambda U_1(x,t) -\lambda^2 J) \psi(x,t,\lambda)=0, \\
M\psi &= i \frac{\partial \psi}{ \partial t } + (V_2(x,t) + \lambda V_1(x,t) -\lambda^2 K) \psi(x,t,\lambda)=0,
\end{split}\end{equation}
where $U_j(x,t)$ and $V_j(x,t)$ are fast decaying smooth functions taking values in the Lie algebra $so(5)$
\begin{equation}\label{eq:2}\begin{aligned}
U_1(x,t) &= [J,Q_1(x,t)], &\quad U_2(x,t) &= [J,Q_2(x,t)] - \frac{1}{2} \ad_{Q_1}^2 J,\\
V_1(x,t) &= [K,Q_1(x,t)], &\quad V_2(x,t) &= [K,Q_2(x,t)] - \frac{1}{2} \ad_{Q_1}^2 K.
\end{aligned}\end{equation}
Here $\ad_{Q_1} X \equiv [Q_1(x,t),X]$ and $J$ and $K$ are two linearly independent constant elements
of the Cartan subalgebra of $so(5)$.

If we assume  $Q_1(x,t)$ and $Q_2(x,t)$ to be generic elements of $so(5)$ then they can be parametrized by:
\begin{equation}\label{eq:3}\begin{aligned}
Q_1(x,t) &= \sum_{\alpha\in\Delta_+ }^{} (q^1_\alpha E_\alpha +p^1_\alpha E_{-\alpha}) + r^1_1H_{e_1} + r^1_2H_{e_2}, \\
Q_2(x,t) &= \sum_{\alpha\in\Delta_+ }^{} (q^2_\alpha E_\alpha +p^2_\alpha E_{-\alpha}) + r^2_1H_{e_1} + r^2_2H_{e_2}, \\
J&=a_1 H_{e_1} + a_2 H_{e_2} =\diag(a_1, a_2, 0 ,-a_2, -a_1) ,  \qquad K=b_1 H_{e_1} + b_2 H_{e_2}=\diag(b_1, b_2, 0 ,-b_2, -b_1),
\end{aligned}\end{equation}
Next we impose on $Q_1(x,t)$ and $Q_2(x,t)$  the natural reduction (\ref{eq:T10.1}a)
\begin{equation}\label{eq:z2}\begin{split}
B_0 U(x,t,\epsilon \lambda^*)^\dag B_0^{-1} = U(x,t,\lambda), \qquad B_0  = \diag( 1, \epsilon, 1,\epsilon, 1), \quad \epsilon^2=1.
\end{split}\end{equation}
As a result:
\begin{equation}\label{eq:chi}\begin{split}
B_0 (\chi^+(x,t,\epsilon\lambda^*))^\dag B_0^{-1} = ( \chi^-(x,t,\lambda))^{-1}, \qquad
B_0 (T(t,\epsilon\lambda^*))^\dag B_0^{-1} = ( T(t,n\lambda))^{-1},
\end{split}\end{equation}
which provide $p^1_\alpha=\epsilon (q^1_\alpha)^*$, $p^2_\alpha=\epsilon (q^2_\alpha)^*$. Then eq. (\ref{eq:nle}) will be
a (rather complicated) system of 8 NLEE for the 8 independent matrix elements $q^1_\alpha$ and $q^2_\alpha$.

However we can impose additional $\bbbz_2$ reduction condition
\begin{equation}\label{eq:inv}\begin{aligned}
D \xi^{\pm} (x,t, - \lambda) \hat{D} &= \xi^\pm(x,t,\lambda), &\qquad
DQ (x,t, -\lambda) \hat{D} &=Q(x,t,\lambda),
\end{aligned}\end{equation}
where  $D=\diag (1,-1,1,-1,1)$.
More precisely this means:
\begin{equation}\label{eq:3'}\begin{aligned}
Q_1(x,t) &= u_1 E_{e_1-e_2} +u_2 E_{e_2} +u_3 E_{e_1+e_2} + v_1 E_{-e_1+e_2} +v_2 E_{-e_2} +v_3 E_{-e_1-e_2} =
\left(\begin{array}{ccccc} 0 & u_1 & 0 & u_3 & 0 \\  v_1 & 0 & u_2 & 0 & u_3 \\  0 & v_2 & 0 & u_2 & 0 \\
  v_3 & 0 & v_2 & 0 & u_1 \\  0 & v_3 & 0 & v_1 & 0  \end{array}\right),\\
Q_2(x,t) &= u_4 E_{e_1} +v_4 E_{-e_1} +w_1 H_{e_1} + w_2 H_{e_2} = \left(\begin{array}{ccccc}  w_1 & 0  & u_4 & 0 &  0 \\
0 & w_2 & 0  &  0 &  0 \\ w_4 & 0  & 0 & 0 & u_4  \\ 0 & 0 & 0 & -w_2 &  0 \\ 0 & 0  & -v_4 & 0 &  -w_1
 \end{array}\right),\\
J&=a_1 H_{e_1} + a_2 H_{e_2} =\diag(a_1, a_2, 0 ,-a_2, -a_1) ,  \qquad K=b_1 H_{e_1} + b_2 H_{e_2}=\diag(b_1, b_2, 0 ,-b_2, -b_1),
\end{aligned}\end{equation}
Combining both reductions for the matrix elements of $Q_j(x,t)$ we have:
\begin{equation}\label{eq:z2a}\begin{split}
v_1 =\epsilon u_1^*, \qquad v_2 =\epsilon u_2^*, \qquad v_3 =\epsilon u_3^*, \qquad v_4 =u_4^*,
\end{split}\end{equation}

The commutativity condition for the Lax pair (\ref{eq:1n})
\begin{equation}\label{eq:L-M}\begin{split}
i \left( \frac{\partial V_2}{ \partial x } +\lambda \frac{\partial V_1}{ \partial x } \right) -
i \left( \frac{\partial U_2}{ \partial t } +\lambda \frac{\partial U_1}{ \partial t } \right)  +
[ U_2 + \lambda U_1  -\lambda^2 J, V_2 + \lambda V_1 -\lambda^2 K]=0
\end{split}\end{equation}
must hold identically with respect to $\lambda$. One can check, that
with the choice (\ref{eq:3}) for $U_i$ and $V_i$, $i=1,2$ the terms proportional to
$\lambda^4$, $\lambda^3$ and $\lambda^2$ vanish identically. The term proportional to $\lambda$ and the $\lambda$-independent
term vanish provided $Q_i$ satisfy the NLEE:
\begin{equation}\label{eq:nle}\begin{split}
& i \frac{\partial V_1}{ \partial x }  -i  \frac{\partial U_1}{ \partial t }  + [U_2,V_1] +[U_1,V_1]=0, \\
& i \frac{\partial V_2}{ \partial x }  -i  \frac{\partial U_2}{ \partial t }  + [U_2,V_2]=0.
\end{split}\end{equation}

In components the corresponding NLEE:
\begin{equation}\label{eq:4-w}\begin{split}
-2i (a_1-a_2) \frac{\partial u_1}{ \partial t } &+2i (b_1-b_2) \frac{\partial u_1}{ \partial x }
+\kappa \epsilon u_2^*(\epsilon u_2^* u_3 - u_1u_2 -2u_4) =0, \\
-2i a_2 \frac{\partial u_2}{ \partial t } &+2i b_2 \frac{\partial u_2}{ \partial x }
-\kappa (u_2\epsilon (|u_3|^2 -|u_1|^2 ) +2 u_3u_4^* +2\epsilon u_1^* u_4) =0, \\
-2i (a_1+a_2) \frac{\partial u_3}{ \partial t } &+2i (b_1+b_2) \frac{\partial u_3}{ \partial x }
+\kappa u_2(\epsilon u_2^* u_3 - u_1u_2 +2u_4) =0,\\
%
-2i a_1 \frac{\partial u_4}{ \partial t } &+2i b_1 \frac{\partial u_4}{ \partial x }
+ i \frac{\partial }{ \partial t } \left( -(2a_2 -a_1) u_1u_2 +(2a_2 +a_1)\epsilon u_2^* u_3 \right)
+ i(2b_2 -b_1) \frac{\partial  (u_1u_2) }{ \partial x } \\
 -i (2b_2 +b_1) \epsilon \frac{\partial  ( u_2^*u_3) }{ \partial x }
& -\kappa \left( 2\epsilon u_4(|u_1|^2 - |u_3|^2)  + \epsilon u_1u_2(|u_1|^2 +3 |u_3|^2) - u_3u_2^*
(3|u_1|^2 + |u_3|^2)\right) =0.
\end{split}\end{equation}

Let us now introduce
\begin{equation}\label{eq:U4}\begin{split}
U_4 = u_4 -\frac{1}{2a_1} ((a_1-a_2) u_1u_2 +(a_1+a_2)\epsilon  u_3u_2^* ).
\end{split}\end{equation}
As a result we get:
\begin{equation}\label{eq:U4eq}\begin{split}
-2i (a_1-a_2) \frac{\partial u_1}{ \partial t } &+2i (b_1-b_2) \frac{\partial u_1}{ \partial x }
-\frac{\kappa \epsilon }{a_1}  u_2^*(2a_1U_4 + \epsilon a_2 u_2^* u_3 +(2a_1-a_2) u_1u_2 ) =0, \\
-2i a_2 \frac{\partial u_2}{ \partial t } &+2i b_2 \frac{\partial u_2}{ \partial x }
-\frac{\kappa \epsilon}{a_1} u_2 ((2a_1+a_2)|u_3|^2 -a_2|u_1|^2 ) - 2\kappa ( u_3U_4^* +\epsilon u_1^* U_4
+u_1^* u_2^*u_3) =0, \\
-2i (a_1+a_2) \frac{\partial u_3}{ \partial t } &+2i (b_1+b_2) \frac{\partial u_3}{ \partial x }
+ \frac{\kappa }{a_1} u_2(\epsilon (2a_1+a_2) u_2^* u_3 - a_2 u_1u_2 +2a_1U_4) =0,\\
-2i a_1 \frac{\partial U_4}{ \partial t } &+2i b_1 \frac{\partial U_4}{ \partial x }
+ \frac{i\kappa}{a_1} \frac{\partial u_1u_2 }{ \partial x }
- \frac{i\kappa \epsilon}{a_1} \frac{\partial u_2^*u_3 }{ \partial x } \\
& -\frac{ \kappa}{a_1} \left( 2\epsilon U_4(|u_1|^2 - |u_3|^2)  + (\epsilon u_1u_2 -  u_3u_2^*)
((2a_1-a_2)|u_1|^2 + (2a_1+a_2) |u_3|^2) \right) =0,
\end{split}\end{equation}

\section{Discussion and conclusions}
On the example of RHP for functions taking values in the $SO(5)$-group we have
outlined the construction of families of commuting operators.
Applied to jets of order 1 with  this method reproduces the
well known results for the $4$-wave equations.

Using jets of order 2 gives us the simplest nontrivial examples for new types of integrable 4-wave equation
whose interaction terms contain quadratic and cubic nonlinearities, as well as $x$-derivatives.
These equations also allow integrable extensions to three-dimensional space-time, see \cite{107s}.

It is not difficult to obtain  other new integrable 4-wave equations.
To this end we can use: i) different power $k$ of the polynomials $U( x ,t,\lambda)$ and $V( x ,t,\lambda)$
  ii) different types of grading; and iii) different reductions of $U$ and $V$.

For example another integrable 4-wave interactions model can be obtained using
Lax pair of degree 4 with respect to $\lambda$:
\begin{equation}\label{eq:1n4}\begin{split}
L\psi &= i \frac{\partial \psi}{ \partial x } + (U_4(x,t) + \lambda U_3(x,t) +\lambda^2 U_2(x,t) + \lambda^3 U_1(x,t)
 -\lambda^4 J) \psi(x,t,\lambda)=0, \\
M\psi &= i \frac{\partial \psi}{ \partial t } + (V_4(x,t) + \lambda V_3(x,t) +\lambda*2
V_2(x,t) + \lambda^3 V_1(x,t) -\lambda^4 K) \psi(x,t,\lambda)=0,
\end{split}\end{equation}
where $U_j(x,t)$ and $V_j(x,t)$ for $j=1,2$ are again as in (\ref{eq:2});
for $j=3$ and $j=4$ we have
\begin{equation}\label{eq:2'}\begin{aligned}
U_3(x,t) &= [J,Q_3(x,t)]  - \frac{1}{2} \left(  \ad_{Q_1} \ad_{Q_2} +  \ad_{Q_2} \ad_{Q_1}\right)J +
\frac{1}{6}  \ad_{Q_1}^3 J , \\
V_3(x,t) &= [K,Q_3(x,t)]  - \frac{1}{2} \left(  \ad_{Q_1} \ad_{Q_2} +  \ad_{Q_2} \ad_{Q_1}\right)K +
\frac{1}{6}  \ad_{Q_1}^3 K , \\
U_4(x,t) &= [J,Q_4(x,t)]  - \frac{1}{2} \sum_{i+j=4}^{} \ad_{Q_i} \ad_{Q_j} J-
\frac{1}{6} \sum_{i+j+k=4}^{} \ad_{Q_i} \ad_{Q_j}\ad_{Q_k} J -\frac{1}{24} \ad_{Q_1}^4 J ,\\
U_4(x,t) &= [K,Q_4(x,t)]  - \frac{1}{2} \sum_{i+j=4}^{} \ad_{Q_i} \ad_{Q_j} K-
\frac{1}{6} \sum_{i+j+k=4}^{} \ad_{Q_i} \ad_{Q_j}\ad_{Q_k} K -\frac{1}{24} \ad_{Q_1}^4 K ,
\end{aligned}\end{equation}

To keep the number of the independent functions down to 4 we will use in addition to
(\ref{eq:T10.1}a), an additional $\bbbz_4$-reduction of the form (\ref{eq:T10.2}) which
leads to the following parametrization of $Q_j$:
\begin{equation}\label{eq:Q1-4}\begin{aligned}
Q_1(x,t) &= u_1 E_{e_1-e_2} +u_2 E_{e_2}+v_1 E_{-e_1+e_2} +v_2 E_{-e_2} , &\qquad Q_2(x,t) &= u_3 E_{e_1-e_2} +v_3 E_{-e_2} , \\
Q_3(x,t) &= = u_4 E_{e_1+e_2} +v_4 E_{-e_1-e_2}, &\qquad Q_4(x,t) &= = w_4 H_{e_1} +z_4 H_{e_2},
\end{aligned}\end{equation}

 It  jets $U( x ,t,\lambda)$  can be view as  elements
of special co-adjoint orbits of the relevant Kac-Moody algebra, generated by the chosen
grading of $\mathfrak{f}$. This allows one to define a hierarchy of Poisson brackets,   see \cite{KuRei,Rei,ReiSTSH},
which along with the conservation laws will provide the hierarchy of Hamiltonian
structures of these NLEE.

Finally we can also  apply Zakharov-Shabat dressing method \cite{ZaSh*74a,ZaSh*79,MiZa*80,RI}
for constructing their explicit ($N$-soliton) solutions.
Instead of solving the inverse scattering problem for $L$ we would rather deal with a Riemann-Hilbert
problem with canonical normalization. For polynomials of order $k$ the contour on which the RHP is
defined consists of $k$ straight lines $l_k \colon \arg \lambda =\pi i/k$ passing through the
origin. Of course, it may necessary to use
dressing factors with more specific $\lambda$-dependence.

This approach can be used also to analyze the NLEE derived by Gel'fand-Dickey approach \cite{GeDi,Harnad}.
It would provide the possibility to systematically construct the spectral decompositions that
linearize the relevant NLEE \cite{IP2,GVYa*08}. Still more challenging is to study the soliton
interactions of the new types of $4$- and $N$-wave equations.

\end{document}